\documentclass[aps,prl,twocolumn,showpacs,superscriptaddress,groupedaddress]{revtex4-1}

\usepackage{graphicx}
\usepackage{dcolumn}
\usepackage{bm}
\usepackage{hyperref,color}

\newcommand{\be}{\begin{equation}}
\newcommand{\ee}{\end{equation}}
\newcommand{\bea}{\begin{eqnarray}}
\newcommand{\eea}{\end{eqnarray}}

\begin{document}

\title{Exact fluctuation theorem without ensemble quantities}

\author{Gregory \surname{Bulnes Cuetara}$^1$}
\author{Massimiliano Esposito$^1$}
\author{Alberto Imparato$^2$}

\affiliation{$^1$ Complex Systems and Statistical Mechanics, University of Luxembourg, L-1511 Luxembourg, Luxembourg \\
$^2$ Department of Physics and Astronomy, Aarhus University, DK-8000 Aarhus C, Denmark}

\date{\today}

\begin{abstract}
Evaluating the entropy production (EP) along a stochastic trajectory requires the knowledge of the system probability distribution, an ensemble quantity notoriously difficult to measure. In this paper, we show that the EP of nonautonomous systems in contact with multiple reservoirs can be expressed solely in terms of physical quantities measurable at the single trajectory level with a suitable preparation of the initial condition. As a result, we identify universal energy and particle fluctuation relations valid for any measurement time. We apply our findings to an electronic junction model which may be used to verify our prediction experimentally.
\end{abstract}


\maketitle
\section{Introduction}

Nowadays experimental techniques enable the measurement of energy and particle fluctuations in very small systems such as single molecules, quantum dots or electric circuits \cite{liphardt2002equilibrium, collin2005verification, douarche2005experimental, Imparato08, fujisawa2006, gustavsson2009, PhysRevX.2.011001, saira2012test, alemany2012experimental, ciliberto2013heat,Imparato13, koski2013distribution}. These fluctuations have been shown to satisfy universal constraints known as fluctuation theorems (FTs) which generalize many former results derived near to equilibrium such as fluctuation-dissipation or Onsager-Casimir reciprocity relations \cite{BochkovKuzovlev1, Evans94, Gallavotti95, Jarzynski97, Kurchan98, crooks1998nonequilibrium, crooks1999entropy, Lebowitz99, seifert2005entropy, andrieux2007fluctuation, esposito2007entropy, RevModPhys.81.1665, campisi2011colloquium, jarzynski2013equalities, Seifert12Rev, EspVDBRev2014}.

Most FTs are nowadays understood as special limiting cases of the finite-time FT for the entropy production (EP) defined at the trajectory level \cite{seifert2005entropy}. For a system in contact with several energy and particle reservoirs this EP becomes the sum of an entropy flow term describing the entropy changes in the reservoirs due to energy and particles currents and a second term describing the change of the system entropy along the trajectory \cite{seifert2005entropy}. Contrary to the entropy flow, this second term is expressed in terms of the initial and final system probability distribution. Since these probabilities are ensemble quantities evaluated at the trajectory level, their experimental measurement has been possible only for systems with very few degrees of freedom \cite{Seifert05exp, Seifert06exp, ciliberto2013heat, koski2013distribution}. Furthermore, even when measurable, these probabilities prevent to express the EP solely in terms of physical observables measurable along a single experimental trajectory. 

Two types of specific setups have been previously considered to resolve this issue. The first one consists of a system driven by a time-dependent force (i.e. nonautonomous) and connected to a single heat reservoir. In this case, the EP reduces to the dissipated mechanical work and the FT reduces to the celebrated Crooks FT \cite{crooks1998nonequilibrium, crooks1999entropy} which has been successfully exploited experimentally \cite{liphardt2002equilibrium, collin2005verification, saira2012test}. The second setup is made of an autonomous system (no time-dependent force) in a nonequilibrium steady state between multiple reservoirs. In the long time limit (in a large deviation sense) the FT for EP reduces to a FT for energy and matter currents. This limit is needed precisely to eliminate the system entropy contribution to the EP. These so-called steady state FTs \cite{esposito2007entropy, andrieux2007fluctuation, Imparato07} have also been verified experimentally \cite{PhysRevB.81.125331, PhysRevX.2.011001}.

\begin{figure}[t]
\centerline{\includegraphics[width=8cm]{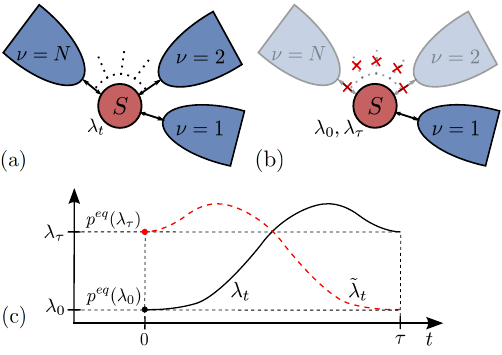}}
\caption{(a) Schematic representation of our setup. (b) Idealized preparation of the initial state of the system, at equilibrium with reservoir $\nu =1$ and disconnected from reservoirs $\nu = 2, \dots , N$. (c) Schematic representation of a forward (full black line) and backward (dashed red line) protocols.}
 \label{figure1}
\end{figure}

In this paper, we show that the EP in general setups can be expressed solely in terms of physical quantities directly measurable at the single trajectory level provided the initial condition of the system is carefully prepared. These setups consist of systems driven by a time-dependent force, in contact with multiple reservoirs, for which the initial conditions correspond to equilibrium with respect to a reference reservoir. We then establish a general finite time FT for the work and the energy and matter currents which reduces to the Crooks FT in presence of a single reservoir and to a finite-time current FT in absence of time-dependent driving. Our FT is particularly relevant for electron counting statistics experiments performed in nonautonomous junctions \cite{saira2012test, koski2013distribution, PhysRevX.2.011001}. 

\section{Finite-time fluctuation theorem}

We consider a system with a discrete set of energy levels denoted by $\epsilon_m(\lambda_t)$ controlled by a time-dependent parameter $\lambda_t$ and connected to $\nu=1, \dots, N$ macroscopic reservoirs with inverse temperatures $\beta_\nu = T_{\nu}^{-1}$ ($k_{\rm B}=1$) and chemical potentials $\mu_\nu$. Our setup is schematically represented in Fig. \ref{figure1} (a).

We start by defining the forward experiment. The system is initially assumed at equilibrium with respect to a reference reservoir denoted $\nu=1$ at the value of the driving parameter $\lambda_0$. We denote its initial probability distribution by $p_{m}^{eq}(\lambda_0)$. Such a state could be prepared by disconnecting all but the reference reservoir and by letting the system relax to equilibrium, see Fig. $\ref{figure1}$ (b). We will later come back to more realistic preparations of such state. All the reservoirs are then simultaneously connected to the system at $t=0$ and the driving parameter $\lambda_t$ starts changing until time $\tau$ where it reaches its final value $\lambda_\tau$. During this time, energy and particles are exchanged between the system and the reservoirs and mechanical work $w_\lambda$ (system energy changes induced by the driving parameter) is performed on the system. After $\tau$, all but the reference reservoir $\nu=1$ are simultaneously disconnected from the system which then reaches after a time $\tau_r$ the equilibrium distribution $p_{m}^{eq} (\lambda_{\tau})$ with respect to the reference reservoir. The entire duration of the forward experiment is $\tau'=\tau+\tau_r$, but we will later see that the disconnection procedure and the subsequent equilibration from $\tau$ to $\tau'$ is in fact not required for our final FT (\ref{generalresult}) to hold.
We denote by ${\bf m}$ the trajectory followed by the system between $0$ and $\tau'$.  
The EP along ${\bf m}$ during this forward experiment reads \cite{esposito2007entropy, PhysRevLett.104.090601}
\be \label{entropygeneral}
\Delta_{{\rm i}} s\left[ {\bf m} |\lambda \right]  =   \ln p^{eq}_{m_0} (\lambda_0) - \ln p^{eq}_{m_{\tau'}} (\lambda_{\tau}) - \sum_{\nu = 1}^{N} \beta_\nu q_\nu \left[ {\bf m}|\lambda \right] ,
\ee
in terms of the equilibrium distributions
\be \label{eqsys}
p_{m}^{eq} (\lambda_t) = \mbox{e}^{- \beta_1 ( \epsilon_m (\lambda_t) - \mu_1 n_m  -\Phi_{1} (\lambda_t) )} , 
\ee
where $\epsilon_{m}(\lambda_t)$ and $n_{m}$ denote the system energy and particle number in state $m$ and $\Phi_{1} (\lambda_t)$ the equilibrium grand potential with respect to $\nu=1$. 
The first two terms in (\ref{entropygeneral}) represent the change in the system entropy. The third one is the entropy change in the reservoirs expressed in terms of the heat $q_\nu \left[ {\bf m}|\lambda \right] \equiv \Delta \epsilon_\nu \left[ {\bf m}|\lambda \right]- \mu_\nu \Delta n_\nu \left[ {\bf m}|\lambda \right]$ where $\Delta \epsilon_\nu \left[ {\bf m}|\lambda \right]$ and $ \Delta n_\nu \left[ {\bf m}|\lambda \right]$ denote respectively the energy and matter flowing from the reservoir $\nu$.

We now consider the backward experiment. The system is initially in the final equilibrium state of the forward protocol $p_{m}^{eq} (\lambda_\tau)$ and all the reservoirs are then reconnected. During a time $\tau$, the system is driven by the time-reversed driving of the forward experiment $\tilde \lambda_t \equiv \lambda_{\tau - t}$ until it reaches its final value $\tilde \lambda_\tau \equiv \lambda_0$. All reservoirs except the reference one $\nu=1$ are then disconnected and the system is allowed to relax to the equilibrium state $p_{m}^{eq} (\tilde \lambda_\tau) = p_{m}^{eq} (\lambda_0)$. Again, we will see that this last step is in fact not required for our FT (\ref{generalresult}) to hold.

A central result in stochastic thermodynamics is that the EP (\ref{entropygeneral}) can be expressed as \cite{crooks1999entropy, seifert2005entropy, esposito2007entropy}
\be
\Delta_{{\rm i}} s\left[ {\bf m} |\lambda \right] = \ln \{P\left[ {\bf m} | \lambda \right] / P [ {\bf \tilde m} | \tilde \lambda ]\},
\label{FT:general}
\ee
where $P\left[ {\bf m} | \lambda \right]$ is the probability to observe a trajectory ${\bf m}$ during the forward experiment and $P [ {\bf \tilde m} | \tilde \lambda ]$ is the probability to observe the time reversed trajectory ${\bf \tilde m}$ during the backward experiment.
As a result, EP satisfies the involution $\Delta_{\rm{i}} s [ {\bf \tilde m} | \tilde{\lambda} ]  = - \Delta_{\rm{i}} s \left[ {\bf m}|\lambda \right]$ under time reversal which directly implies the FT \cite{seifert2005entropy, PhysRevLett.104.090601, Verley_2012_PhysicalReviewE} 
\be
\ln \frac{P(\Delta_{{\rm i}} s)}{\tilde{P}( - \Delta_{{\rm i}} s)} = \Delta_{{\rm i}} s ,
\ee
where $P(x)$ and $ \tilde{P}(x)$ are the probability distributions of the EP during the forward and backward experiment, respectively. 

We now make use of energy and particle number conservation at the single trajectory level. Energy changes are separated into contributions due to the driving parameter (mechanical work) and the reservoirs (energy flows) while particle changes are only due to particle flows 
\bea
 && \epsilon_{m_{\tau'}}(\lambda_{\tau'}) - \epsilon_{m_0} (\lambda_0) = w_{\lambda} \left[ {\bf m}|\lambda \right] + \sum_{\nu = 1}^{N} \Delta \epsilon_\nu \left[ {\bf m}| \lambda \right]\label{energyconservation} \\
&& n_{m_{\tau'}}(\lambda_{\tau'} ) - n_{m_0} (\lambda_0) = \sum_{\nu = 1}^{N} \Delta n_\nu \left[ {\bf m}|\lambda \right] \label{particleconservation}.
\eea
Together with the equilibrium condition (\ref{eqsys}), we can rewrite the EP as 
\be \label{entropyinitialcond}
\Delta_{{\rm i}} s   = 
\beta_1 \left( w_\lambda - \Delta \Phi_1 \right) + \sum_{\nu = 2}^{N} \left( A^{\epsilon}_\nu \Delta \epsilon_\nu + A^{n}_\nu \Delta n_\nu \right) ,
\ee
where we omitted the trajectory dependence to lighten the notation and introduced the thermodynamic forces 
\bea
A^{\epsilon}_\nu = \beta_1-\beta_\nu \ \ \;, \  \ A^{n}_\nu = \beta_\nu  \mu_\nu - \beta_1 \mu_1 ,
\eea
associated to the energy and matter transfers respectively.
We also defined the change in the reference grand potential $\Delta \Phi_1 = \Phi_{1} (\tau) - \Phi_{1} (0)$ which only depends on the initial and final values of the driving parameter and thus does not fluctuate. As a result, the FT for the EP can be written as
\begin{widetext}
\begin{eqnarray}
\ln \frac{P( \beta_1 w_\lambda + \tau \sum_{\nu = 2}^{N} \left[ A^{\epsilon}_\nu j_{\nu}^{\epsilon } + A^{n}_\nu j_{\nu}^{n} \right])}{\tilde P( - \beta_1 w_{\lambda} -\tau \sum_{\nu = 2}^{N} \left[ A^{\epsilon}_\nu  j_{\nu}^{\epsilon } + A^{n}_\nu j_{\nu}^{n } \right] )} 
= \ln \frac{P(w_\lambda , \{j_{\nu}^{\epsilon }\} , \{j_{\nu}^{n}\})}{\tilde P( -w_{\lambda} , \{-j_{\nu}^{\epsilon }\},\{-j_{\nu}^{n }\})} 
= \beta_1 \left( w_\lambda - \Delta \Phi_1 \right) + \tau \sum_{\nu=2}^{N} \left( A^{\epsilon}_\nu j_{\nu}^{\epsilon } + A^{n}_\nu j_{\nu}^{n } \right) ,
\label{generalresult}
\end{eqnarray} 
\end{widetext}
in terms of the energy and particle currents, $j_{\nu}^{\epsilon } = \Delta \epsilon_\nu / \tau$ and $j_{\nu}^{n }= \Delta n_\nu / \tau$, entering the system from reservoir $\nu$.
The first equality in (\ref{generalresult}) results from the fact that when EP is a sum of odd terms under time reversal, a detailed FT also holds for their joined probability distribution \cite{PhysRevE.82.030104}. The FT (\ref{generalresult}) is our main result. It holds for specific initial conditions corresponding to equilibrium with respect to a reference reservoir (\ref{eqsys}), but is valid for any time and is exclusively expressed in terms of physical observables at the trajectory level: the mechanical work performed on the system $w_\lambda$ and the energy and particle currents $j_{\nu}^{\epsilon}$ and $j_{\nu}^{n}$. The generalization to time dependent temperatures and chemical potentials is given in the Appendix. 

As previously announced, disconnecting all but the reference reservoir and letting the system relax in the forward as well as in the backward experiment is in fact not needed. All the fluctuating quantities appearing in the argument of the probability distributions in (\ref{generalresult}) stop evolving during these relaxation process, i.e. $w_\lambda = j_{\nu}^{\epsilon} = j_{\nu}^{n} = 0$ for $\nu=2,\dots , N$. As a result, the measurement of the mechanical work and of the fluxes can be performed during any chosen time regardless of the final state of the system and of the type of driving.

The initial condition (\ref{eqsys}) can be prepared without disconnecting the reservoirs, by letting the system relax with all temperatures and chemical potentials of the reservoirs set to $\beta_1$ and $\mu_1$. These latter are then simultaneously switched to their nominal values $\beta_\nu$ and $\mu_\nu$ on a time scale shorter than that of the system dynamics. 
We show in the Appendix that such switching is not affecting (\ref{generalresult}). Alternatively, one may directly weight measurement outcomes along single trajectories with the equilibrium distribution (\ref{eqsys}) in experimental situations for which the initial state of the system is controlled.

In presence of a single reservoir, $A^{\epsilon}_\nu = A^{n}_\nu = 0$, (\ref{generalresult}) reduces to the Crooks FT \cite{crooks1998nonequilibrium, crooks1999entropy} for the mechanical work $\ln \{P\left[ {\bf m} | \lambda \right] / P [ {\bf \tilde m} | \tilde \lambda ]\} = \exp \left[\beta \left( w_\lambda \left[ {\bf m} | \lambda \right] - \Delta \Phi \right)\right]$. 
On the other hand, in absence of time-dependent driving our FT (\ref{generalresult}) becomes equivalent to a current FT \cite{esposito2007entropy, andrieux2007fluctuation, RevModPhys.81.1665} 
\be
\ln  \frac{P( \{j_{\nu}^{\epsilon }\}, \{j_{\nu}^{n}\} ) }{ P( \{ - j_{\nu}^{\epsilon}\}, \{- j_{\nu}^{n}\})} = \tau \sum_{\nu=2}^{N} \left( A^{\epsilon}_\nu j_{\nu}^{\epsilon } +A^{n}_\nu j_{\nu}^{n } \right) .
\ee 
Remarkably this FT is now valid at all times \cite{gregth} (due to our choice of initial condition) while when initially at steady state a long time limit is needed.

For isothermal setups ($\beta_{\nu}=\beta$ for all $\nu$), our FT (\ref{generalresult}) simplifies to
\be \label{chemworkFT}
\hspace{-0.2cm} \ln \frac{P(w_\lambda+w_{c} )}{ \tilde P(-w_{ \lambda}- w_{c} )} 
= \ln \frac{P(w_\lambda , w_{c} )}{ \tilde P(- w_{ \lambda} , - w_{c} )} 
= \beta \left( w_\lambda +  w_{c} - \Delta \Phi_1 \right) ,
\ee
where the chemical work for transferring particles from one reservoir to another is
\be
w_{c} = \tau \sum_{\nu = 2}^{N} \left(\mu_\nu - \mu_1 \right) j_{\nu}^{n }.
\ee
Despite their very different nature, mechanical and chemical work play the same role in this result. 

\section{Model system}

As a concrete application, we consider a single level quantum dot connected to two electronic reservoirs at equilibrium with the same temperature but different chemical potentials. We also assume that an external field drives the energy of the single level as $\epsilon_t= \epsilon - a \cos \omega t$ with an amplitude $a$ and frequency $\omega$. The corresponding backward protocol for a given measurement time $\tau$ is $\tilde{\epsilon}_t \equiv  \epsilon_{\tau -t}$. If $m=0,1$ denote respectively the empty and filled single level, in the Coulomb blockade regime the dynamics of the occupation probabilities $p_{m}$ is described by the master equation $\dot p_m = \sum_{m, m'} \Gamma_{mm'} p_{m'}$ where $\Gamma_{mm}=-\sum_{m'} \Gamma_{m'm}$ \cite{RevModPhys.81.1665, PhysRevB.74.235309}. The Fermi's Golden Rule rate to charge and uncharge the dot are respectively given by $\Gamma_{10} = \sum_{\nu = 1,2} \gamma_{\nu} f_{\nu} (\epsilon_t)$ and $\Gamma_{01} = \sum_{\nu = 1,2} \gamma_{\nu} (1 -f_{\nu} (\epsilon_t))$ in terms of the tunneling rates $\gamma_{\nu}$ and the Fermi-Dirac distribution function $f_{\nu}(x) = (1+\exp \beta(x-\mu_{\nu}))^{-1}$ of the reservoirs $\nu = 1, 2$.
Similar experimental setups have been considered in Refs. \cite{saira2012test, PhysRevX.2.011001}. Since the single electron transfers and the dot occupation can be monitored experimentally, these setups are ideal to verify our predictions.

For this isothermal setup, we will explicitly verify the FT (\ref{chemworkFT}) by numerically calculating the statistics of the mechanical and chemical work using the generating function techniques developed in Ref. \cite{esposito2007entropy}. The mechanical work $w_\lambda$ is the energy provided by the external field to lift the energy of the single level when charged. When uncharged, no mechanical work is performed by the field. The chemical work $w_c =\tau \Delta \mu j^{n}_2$, with $\Delta \mu = \mu_2 - \mu_1$, is in turn the energy needed to transfer $\tau j^{n}_2$ electrons from reservoir $\nu=2$ to reservoir $\nu=1$.

\begin{figure}[h]
\centerline{\includegraphics[width=8.7cm]{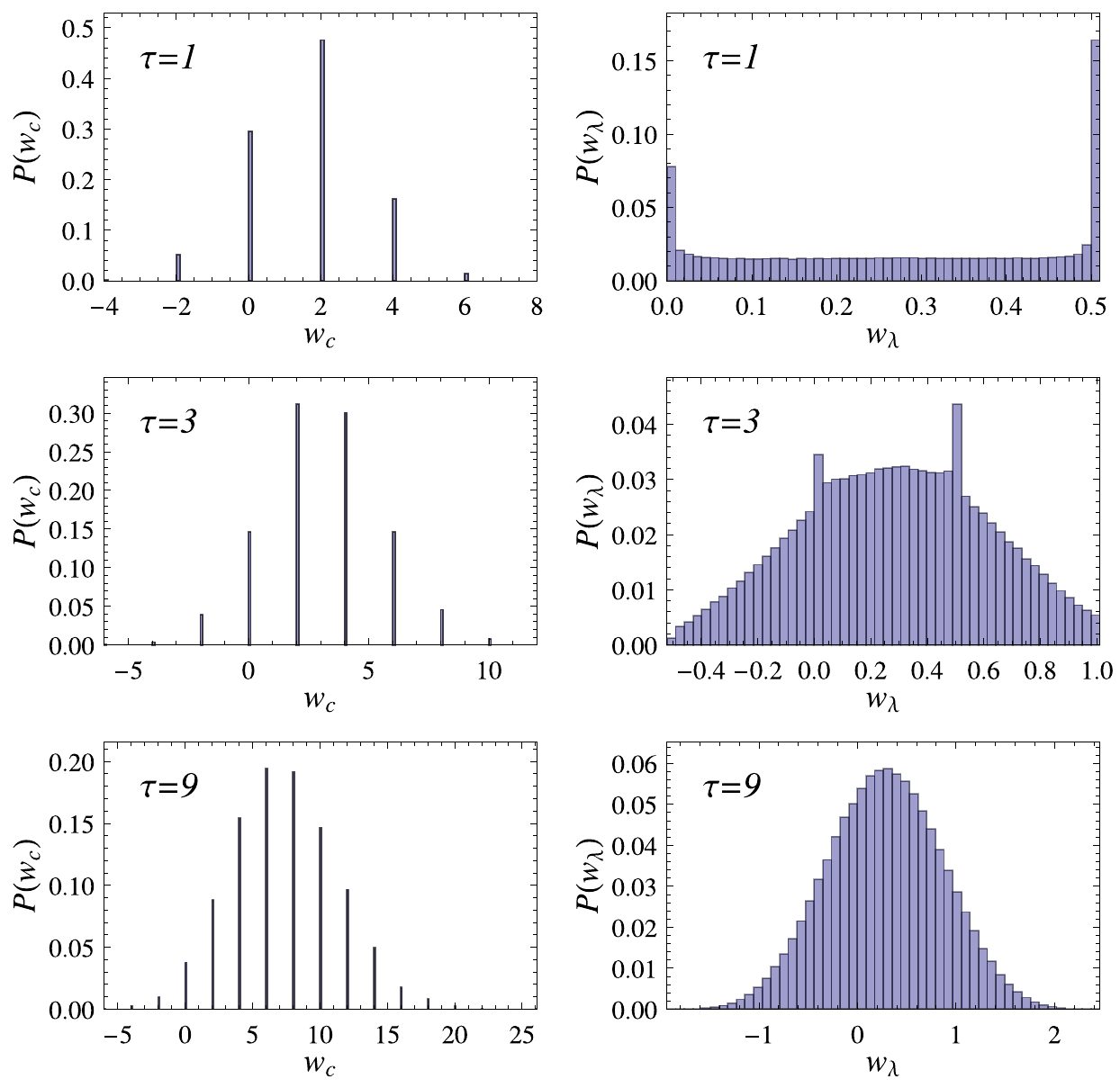} }
\caption{Simulations of the chemical work distribution $P(w_c)$   (left column) and of the mechanical work distribution $P(w_\lambda) dw_\lambda \equiv P(w_\lambda \in \left[ w_\lambda , w_\lambda + dw_\lambda\right])$ (right column), along the forward protocol and for three different measurement times $\tau$. Each histogram contains $50$ bins which span the support of the distribution. We used $\beta = 1$, $\mu_1 = 1$, $\mu_2 = 3$, $\epsilon = 2$, $a=0.5$, $\omega = \pi /2$, $\gamma_1 = 1.5$, and $\gamma_2 = 1.4$. }
\label{distrhisto}
\end{figure}

Simulations of the chemical work distribution $P(w_c)$ are illustrated on the left column of Fig. \ref{distrhisto} for several values of the measurement time $\tau$. The chemical work takes discrete values $w_c = k \Delta \mu$ where $k$ is the number of particles transferred from reservoir $2$ to $1$ during time $\tau$. The distribution spreads and drifts as the measurement time $\tau$ increases. 

Simulations of the mechanical work distribution $P(w_\lambda)$ are depicted on the right column of Fig. \ref{distrhisto}. At short times, i.e. for $\tau \ll \Gamma^{-1}_{10} ,\Gamma^{-1}_{01} $, electron transfers barely occur during the measurement and the distribution is essentially $P(w_\lambda) \sim p^{eq}_{0} \delta (w_\lambda)+ p^{eq}_{1} \delta \left(w_\lambda- a \cos (\omega \tau)  \right)$. It becomes smoother as electrons begin to dwell randomly in the quantum dot for increasing measurement times. The initial peaks completely disappear for $\tau \gg \Gamma^{-1}_{10} ,\Gamma^{-1}_{01}$. This distribution has a limited support determined by the minimum and maximum work that can be done by the protocol on the quantum dot during $\tau$. 

\begin{figure}[h] 
\centerline{\includegraphics[width=8.7cm]{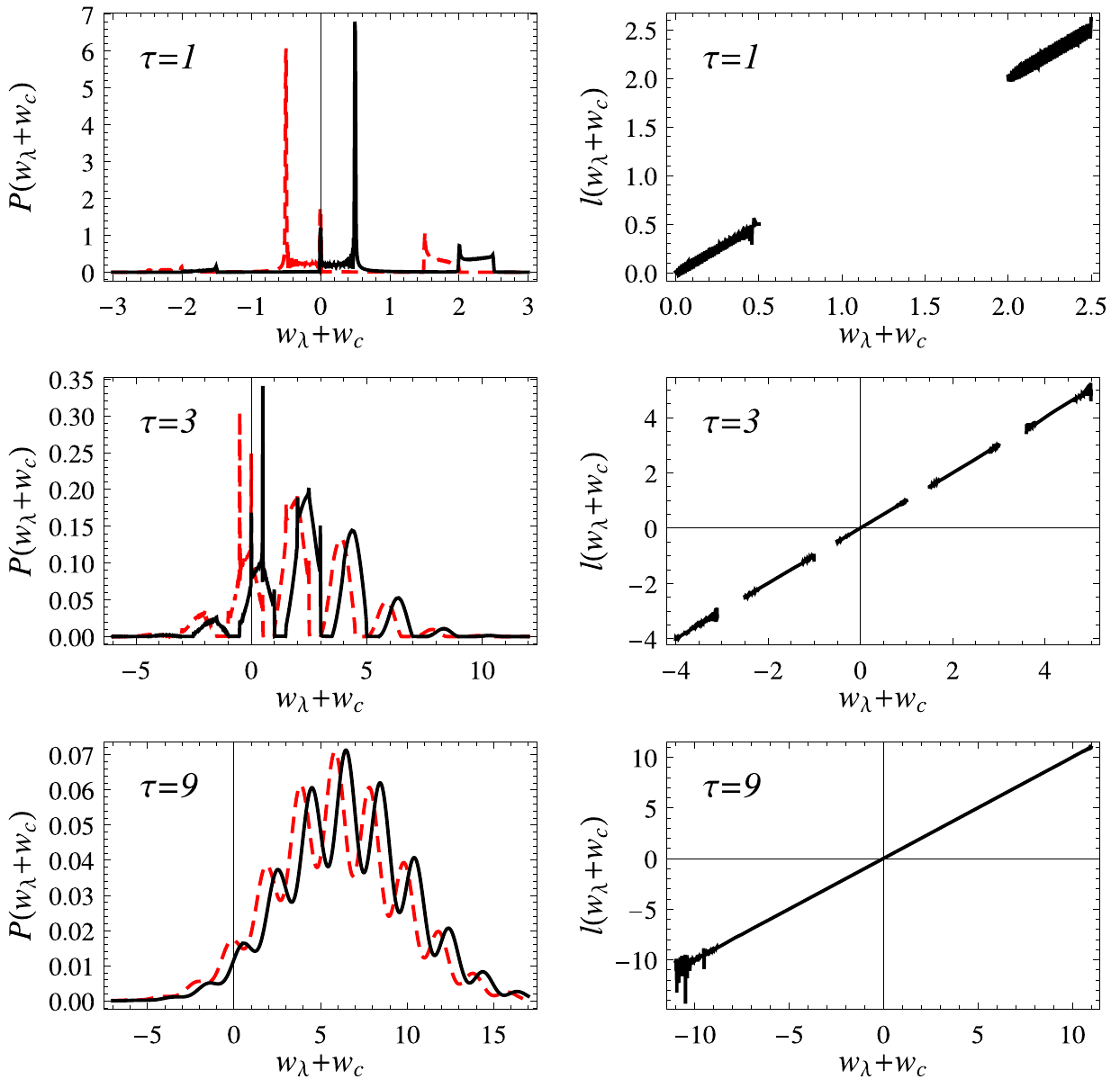} }
\caption{Left column: Probability distribution for the sum of the mechanical and chemical work along the forward (full black lines) and backward (dashed red lines) protocols for different measurement times $\tau$. Right column: Explicit verification of the fluctuation relation (\ref{chemworkFT}). The quantity on the ordinate is $l(w_\lambda + w_c)$ $\equiv \ln [P(w_\lambda + w_c)/\tilde P(-w_{\lambda} - w_c)] + \beta \Delta \Phi$ and equals $w_\lambda + w_c$ when the FT is satisfied. Parameters are the same as in Fig. \ref{distrhisto}.
}
\label{fullandtest2}
\end{figure}

The distributions for the mechanical and chemical works, $P(w_c)$ and $P(w_\lambda)$, in general do not satisfy a FT for finite time $\tau$. However, their joined distribution $P(w_c,w_\lambda)$ as well as the distribution for their sum $P(w_c+w_\lambda)$ do satisfy the FT (\ref{chemworkFT}). Numerical evaluations of this latter along the forward and backward protocol by use of the generating function techniques \cite{esposito2007entropy}  are shown in the left column of Fig. \ref{fullandtest2}. The oscillations in these distributions can be understood by noting that $P(w_\lambda + w_c)$ $= \sum_{k} P (w_\lambda - k \Delta \mu , k \Delta \mu)$. Provided the width of the mechanical work distribution is of order $\Delta \mu$ or smaller, oscillations are to be expected. Moreover, for short measurement times, the distribution is identically zero on finite subsets of the work axis due to the limited amount of work that can be performed by the mechanical driving. 
The FT (\ref{chemworkFT}) is explicitly verified in the right column of Fig. \ref{fullandtest2}. Here again, portions of work are missing at short times due to the limited support of the mechanical work distribution.

\section{Conclusion}

Various fluctuation relations have been derived in the recent years many of which lack a direct connection to experimental observables. The FT derived in this paper is solely expressed in terms of physical observable at the trajectory level and generalizes the former experimentally relevant FTs to setups involving time dependent forces and multiple reservoirs. 

\begin{acknowledgments}

\section{Acknowledgments}

G.B.C. and M.E. are supported by the National Research Fund, Luxembourg in the frame of project FNR/A11/02 and A.I. by the Lundbeck Fonden and by The Danish Council for Independent Research-- Natural Sciences.

\end{acknowledgments}

\appendix

\section{Appendix}

We now consider the generalization of the FT (\ref{generalresult}) in presence of time dependent inverse temperatures $\beta_{\nu} (t)$ and chemical potentials $\mu_{\nu} (t)$ for $\nu = 1 , \dots , N$.

The system is assumed to be described by a stochastic master equation whose transition rates satisfy the local detail balance condition
\bea \label{kmssupp}
\ln \frac{\Gamma_{m m'}^{\nu}(t)}{\Gamma_{m' m}^{\nu} (t)}
= -\beta_\nu (t) \left( \Delta \epsilon_{\nu} - \mu_\nu (t) \Delta n_\nu \right) ,\label{ldb}
\eea
where $\Delta \epsilon_{\nu}$ and $\Delta n_{\nu}$ denote respectively the amount of energy and particles flowing out of reservoir $\nu$ during the transition from $m'$ to $m$. 

A system trajectory ${\bf m}$ is a particular realisation of the stochastic process during which the system undergoes a succession of transitions at times $t_i$ for $i = 1, \dots , k$ involving an energy and particle number exchange $\Delta \epsilon_{\nu}^{i}$ and $\Delta n_{\nu}^{i}$ with a given reservoir $\nu$. In the following, we make the identifications $t_0 = 0$ and $t_{k +1} = \tau$, and denote by $\epsilon_{m_i}(t)$ and $n_{m_i}$ the energy and particle number of the system in state $m_i$ at time $t$. 

We introduce the instantaneous energy and matter currents out of reservoir $\nu$, and the mechanical power respectively as
\bea \label{curr}
j_{\nu}^{\epsilon} (t) & \equiv & \tau^{-1} \sum_{i=1}^{k} \Delta \epsilon^{i}_{\nu} \,\delta (t - t_i) \\
j_{\nu}^{n} (t) & \equiv & \tau^{-1} \sum_{i=1}^{k} \Delta n^{i}_{\nu} \, \delta (t - t_i)  \\
\dot w_\lambda (t) & \equiv & \sum_{i=0}^{k} \dot \epsilon_{m_{i}} (t) \, \chi_i (t) \label{pow}
\eea
in terms of the Dirac delta function $\delta (t)$ and step functions $\chi_i (t)$ which are equal to $1$ for $t \in ] t_i , t_{i+1} [$ and $0$ otherwise. The conservation laws (\ref{energyconservation}) - (\ref{particleconservation}) are then equivalent to the constrains
\bea
\dot \epsilon (t) & = & \dot w_\lambda (t) + \sum_{\nu = 1}^{N} j_{\nu}^{\epsilon} (t) \label{singtimecons1} \\
 \dot n (t) & = &  \sum_{\nu = 1}^{N} j_{\nu}^{n} (t) \label{singtimecons2}
\eea
in terms of the energy $\epsilon (t)$ and number of particles $n(t)$ in the open system at time $t$.

Using the expression (\ref{FT:general}) for the EP as well as (\ref{kmssupp})-(\ref{singtimecons2}), we find that
\bea 
&&\Delta_{\rm i} s \left[ {\bf m} | \{ \beta_{\nu}(t) \} , \{ \mu_{\nu}(t) \} , \lambda (t) \right] \label{EPtimedepchem} \\
&&\hspace{0.3cm}= \ln p_{m_{0}} - \ln p_{m_{\tau}} + \sum_{i=1}^{k} \ln \frac{\Gamma_{m_{i-1} m_i} \left(  t_i  \right)}{\Gamma_{m_{i} m_{i-1}} \left(  t_i  \right)} \nonumber 
\eea
\bea
&&\hspace{0.3cm}= \left[ \beta_{1} (t) \left( \epsilon_{m_t} (t) - \mu_{1} (t) n_{m_t} \right) \right]_{t=0}^{t=\tau} \nonumber \\
&&\hspace{0.8cm}- \int_{0}^{\tau} dt \, \sum_{\nu = 1}^{N} \beta_{\nu} (t) \left( j_{\nu}^{\epsilon} (t) - \mu_{\nu} (t) j_{\nu}^{n} (t) \right) \nonumber \\
&&\hspace{0.3cm}= \int_{0}^{\tau} dt \, \left( \dot \beta_1 (t)  \epsilon (t) ) - \frac{d}{dt}{\left( \beta_1 (t) \mu_1 (t) \right)} n(t) \right) \nonumber \\
&&\hspace{0.8cm}+ \int_{0}^{\tau} dt \, \sum_{\nu = 2}^{N} \left( A_{\nu}^{\epsilon} (t) j_{\nu}^{\epsilon} (t) + A_{\nu}^{n} (t) j_{\nu}^{n} (t) \right) \nonumber \\
&&\hspace{0.8cm}- \beta_1 (\tau) \phi_1 (\tau) + \beta_1 (0) \phi_1 (0) + \int_{0}^{\tau} dt \, \beta_1 (t) \dot w_\lambda (t). \nonumber 
\eea
Assuming that the inverse temperature and chemical potential of the reference reservoir $\nu = 1$ is time-independent, we get
\bea
\ln \frac{P (\Delta_{\rm i} s)}{ \tilde{P}(-\Delta_{\rm i} s)} &=&  \beta_1  (w_\lambda  -  \Delta \phi_1  ) \label{generalfttimedepchem} \\
&&+ \int_{0}^{\tau} dt \, \sum_{\nu = 2}^{N} \left( A_{\nu}^{\epsilon} (t) j_{\nu}^{\epsilon} (t) + A_{\nu}^{n} (t) j_{\nu}^{n} (t) \right),\nonumber
\eea
expressed in terms of the time dependent thermodynamic forces
\bea
&A^{\epsilon}_\nu (t) = \beta_1 -\beta_\nu (t) \\
&A^{n}_\nu (t) = \beta_\nu (t)  \mu_\nu (t) - \beta_1  \mu_1  ,
\eea
and the grand canonical potential difference 
\bea
\Delta \phi_1 = \phi_1 (\tau) - \phi_1 (0).
\eea

The FT (\ref{generalfttimedepchem}) is the generalization of (\ref{generalresult}) when considering time dependent temperatures and chemical potentials in all the reservoirs. 

As announced in the paper, a sudden switch in the temperatures and chemical potentials of all but the reference reservoir does not contribute to the EP appearing in the right-hand side of the FT (\ref{generalfttimedepchem}). Indeed, since the switch is performed on a time scale shorter than the typical time scale of transfer processes between the reservoirs and the system, all the currents remain zero during the switch $j^{\epsilon}_\nu (t_s) = j^{n}_{\nu} (t_s) =w_\lambda  =  \Delta \phi_1 = 0$. We implicitly assumed that the relaxation time scale of the reservoirs is much shorter than all other relevant time scale.

%

\end{document}